\documentstyle[12pt]{article}
\def\ref{\par\hangindent=1.0cm\hangafter=1}

\begin{document}

\baselineskip=20pt plus 1pt minus 1pt
\textwidth=6.5truein
\textheight=9.0truein

\begin{center}
{\bf Coupled Q-oscillators as a model for}
{\bf  vibrations of polyatomic molecules}
\vskip 0.2in

Dennis Bonatsos $^{+ *}$ \footnote[1]{e-mail: bonat@ectstar.ect.unitn-it, 
bonat@cyclades.nrcps.ariadne-t.gr},
 C. Daskaloyannis $^\dagger$ \footnote[2]{e-mail:
daskaloyanni@olymp.ccf.auth.gr} and P. Kolokotronis $^+$\\
\vskip 0.1  in

$^+$ Institute of Nuclear Physics, NCSR ``Demokritos''\\
GR-15310 Aghia Paraskevi, Attiki, Greece\\
$^*$ European Centre for Theoretical Studies in Nuclear Physics and Related 
Areas (ECT$^*$)\\
Villa Tambosi, Strada delle Tabarelle 286, I-38050 Villazzano (Trento), 
Italy\\  
$^\dagger$ Theoretical Physics Department, Aristotle University of 
Thessaloniki\\
GR-54006 Thessaloniki, Greece

\vskip 0.2in
{\bf Abstract}
\end{center}

The system of two $Q$-deformed oscillators coupled so that the total
Hamiltonian has the su$_Q$(2) symmetry is proved to be equivalent,
to lowest order approximation, to a system of two identical Morse
oscillators
coupled by the cross-anharmonicity usually used empirically in describing
vibrational spectra of triatomic molecules. The symmetry also
imposes a connection between the self-anharmonicity of the Morse oscillators
and the cross-anharmonicity strength, which can be removed by replacing
the $Q$-oscillators by deformed anharmonic oscillators.
The generalization to $n$ oscillators is straightforward. The
applicability of the formalism to highly symmetric polyatomic
molecules is discussed.

\newpage

Quantum algebras (also called quantum groups) $^{1,2}$,
are recently receiving much attention in physics. They are
$q$-deformations
of the universal enveloping algebras  of the corresponding Lie
algebras, to which they reduce when the deformation parameter $q$
is set equal to 1. From the mathematical point of view they are Hopf 
algebras $^3$. 
Initially used for solving the quantum Yang Baxter
equation $^4$, they are now finding applications in several
branches of physics, especially after the introduction of the
$q$-deformed harmonic oscillator $^{5-7}$. Applications in
conformal field theory, quantum gravity,
quantum optics, atomic physics, nuclear physics, 
as well as in the description of spin chains have already appeared.

In molecular physics rotational spectra of diatomic molecules have
been described in terms of the su$_q$(2) symmetry $^{8,9}$.
For vibrational spectra of diatomic molecules both $q$-deformed
harmonic $^{10}$ and $q$-deformed anharmonic $^{11}$
oscillators have been successfully used and WKB equivalent potentials
giving the same spectrum as these oscillators have been determined
$^{12}$, related to a
$q$-deformation of the modified P\"oschl Teller potential,  which is
connected to the Morse potential by a known transformation $^{13}$.
A review of applications of quantum algebraic techniques in diatomic 
molecules has been given in $^{14}$. 

The success of the quantum algebraic description in vibrational
spectra of diatomic molecules
creates the question if an extension to vibrational spectra of
triatomic and polyatomic
molecules is possible. Vibrational spectra of triatomic and polyatomic
molecules have been traditionally described in terms of coupled
oscillators $^{15-19}$. Recently, a model of coupled
anharmonic oscillators, in which each bond in a polyatomic molecule
is replaced by a Morse potential, has also appeared $^{20-23}$.
(This model is a simplification of the vibron model $^{24,25}$,
in which both rotations and vibrations are treated simultaneously.)
On the other hand, the way of coupling $n$ $q$-deformed oscillators
so that the total Hamiltonian is characterized by an su$_q$(n)
symmetry has been understood $^{26-28}$. The spectrum of this
system exhibits cross-anharmonicities among the levels of the individual
oscillators, imposed by the quantum symmetry. It is therefore
worth examining if the cross-anharmonicities imposed by the overall
quantum symmetry bear any similarity to the cross-anharmonicities long used
in molecular physics on an empirical basis.

The Hamiltonian usually used for the description of vibrational
modes of polyatomic molecules is $^{15}$
$$ H = \sum_i \omega_i \left(v_i+\frac{d_i}{2}\right) +\sum_i \sum_{k\geq i}
x_{ik} \left(v_i+\frac{d_i}{2}\right)\left(u_k+\frac{d_k}{2}\right), \eqno(1)$$
where $v_i$, $v_k$ are vibrational quantum numbers and $d_i$,
$d_k$ the degeneracies of the corresponding modes. More
specifically, for ABA triatomics a frequently used Hamiltonian
reads $^{17-19}$
$$H= \hbar \omega_1 \left(n_1+\frac{1}{2}\right) +\hbar \omega_2
\left(n_2+\frac{1}{2}\right) $$ $$+ \frac{\gamma_1}{2} 
\left(n_1+\frac{1}{2}\right)^2
+\frac{\gamma_2}{2} \left(n_2+\frac{1}{2}\right)^2 +\gamma_{12}
\left(n_1+\frac{1}{2}\right) \left(n_2+\frac{1}{2}\right). \eqno(2)$$
It is clear that the first and third term in this Hamiltonian describe
an anharmonic oscillator, the second and fourth term describe another
anharmonic oscillator, while the fifth term describes the cross-anharmonicity
between them.  It is  worth recalling at this point that the spectrum
of the anharmonic oscillators encountered here is the same as the
spectrum obtained from solving the Schr\"odinger equation for the
Morse potential $^{29}$.

In recent work $q$-numbers are defined as
$$ [x]_q = \frac{q^x-q^{-x}}{q-q^{-1}}, \eqno(3)$$
where $q$ can be real ($q=e^{\tau}$, where $\tau$ real) or a
phase factor ($q=e^{i\tau}$, with $\tau$ real).
The $q$-deformed harmonic oscillator $^{5-7}$ is defined in terms
of the creation and annihilation operators $a^\dagger$ and $a$ and the
number operator $N$, which satisfy the commutation relations
$$ [N, a^\dagger] = a^\dagger, \qquad [N, a]=-a, \qquad
a a^\dagger - q^{\mp} a^\dagger a = q^{\pm N}. \eqno(4)$$
The Hamiltonian of the $q$-deformed harmonic oscillator is
$$H= \frac{\hbar\omega}{2} (a a^\dagger + a^\dagger a), \eqno(5)$$
and its eigenvalues are
$$E(n)= \frac{\hbar\omega}{2} ([n]_q + [n+1]_q). \eqno(6)$$

Coupling two of these oscillators with the requirement that the total
Hamiltonian has the su$_q$(2) symmetry one finds $^{26-28}$
$$H = a_1^\dagger a_1 q^{N_2} + q^{-N_1} a_2^\dagger a_2, \eqno(7)$$
the generators of su$_q$(2) in terms of two mutually commuting
sets of $q$-boson operators ($a_1$, $a_1^\dagger$, $N_1$) and ($a_2$, 
$a_2^\dagger$, $N_2$) in the Schwinger realization being
$$ J_0=\frac{1}{2} (N_1-N_2), \qquad J_+= a_1^\dagger a_2, \qquad
J_-=a_2^\dagger a_1 , \eqno(8)$$
and satisfying the commutation relations
$$ [J_0, J_{\pm}] =\pm  J_{\pm}, \qquad [J_+, J_-]=[2 J_0]_q.
\eqno(9)$$
The relevant eigenvalues of the Hamiltonian of Eq. (7) are
$$E(n_1,n_2) = [n_1+n_2]_q. \eqno(10)$$
For $l$ q-deformed harmonic oscillators coupled so that the overall
symmetry is su($l$) one obtains the eigenvalues $^{26-28}$
$$E(n_1,n_2, \dots, n_l) = [n_1+n_2+ \dots +n_l]_q. \eqno(11)$$
Using the Taylor expansion
$$[N]_q= N \pm \frac{\tau^2}{6} (N-N^3) + \frac{\tau^4}{360}
(7N-10N^3+3N^5) \pm \cdots, \eqno(12)$$
where the upper (lower) sign corresponds to $q$ being a phase factor (real),
the energy eigenvalues of Eq. (10), including terms up to $\tau^2$,
can be written as
$$E(n_1,n_2) = n_1 \left(1\pm\frac{\tau^2}{6}\right) \mp \frac{\tau^2}{6}
 n_1^3 + n_2 \left(1\pm \frac{\tau^2}{6}\right) \mp \frac{\tau^2}{6}
n_2^3 $$ $$\mp \frac{\tau^2}{2} (n_1^2 n_2 + n_2^2 n_1) + \dots \eqno(13)$$
In comparison to the empirical Hamiltonians of Eqs (1) and (2) we
remark that :

i) The lowest order
self-anharmonicities of the oscillators involved in Eq. (13) are
proportional to $n_i^3$ and not to $n_i^2$ as in Eqs (1), (2).

ii) The lowest order cross-anharmonicities in Eq. (13) are of the type
$n_i^2 n_j$ and $n_j^2 n_i$, and not of the type $n_i n_j$ as in
Eqs (1), (2).

iii) The previous two remarks also hold in the case of $n$ coupled
$q$-deformed oscillators.

We therefore conclude that the usual coupled $q$-deformed oscillators
are not suitable for the description of vibrational spectra of
polyatomic molecules.

However, a different version of deformed oscillator, the $Q$-oscillator
$^{30-32}$, can be simply obtained by defining $^{32,33}$
the operators $b$, $b^\dagger$ through the equations
$$a = q^{1/2} b q^{-N/2}, \qquad a^\dagger = q^{1/2} q^{-N/2} b^\dagger.
\eqno(14)$$
Eq. (4) then gives
$$ [N,b^\dagger]= b^\dagger, \qquad  [N,b]=-b, \qquad 
b b^\dagger- q^2 b^\dagger b =1.\eqno(15)$$
By using the symbol $Q=q^2$ and introducing the Q-number
$$[x]_Q= \frac{Q^x-1}{Q-1}, \eqno(16)$$
where $Q=e^T$ (with $T$ real),
we have that
$$ b^\dagger b = [N]_Q, \qquad bb^\dagger = [N+1]_Q. \eqno(17)$$
The basis is defined by
$$b |0>=0, \qquad |n>={(b^\dagger)^n \over \sqrt{[n]_Q!} } |0>,\eqno(18) $$
where the $Q$-factorial is defined by
$$ [n]_Q! = [n]_Q [n-1]_Q \dots [2]_Q [1]_Q. \eqno(19)$$
The action of the operators on the basis is given by
$$ N |n> = n |n>, \qquad b^\dagger |n> = \sqrt{[n+1]_Q} |n+1>, \qquad 
b |n>= \sqrt{ [n]_Q} |n-1>.\eqno(20)$$
The Hamiltonian of the corresponding deformed  harmonic oscillator
has the form
$$H =\frac{\hbar\omega}{2} (b b^\dagger + b^\dagger b), \eqno(21)$$
the eigenvalues of which are
$$E(n) =\frac{\hbar\omega}{2} ([n]_Q +[n+1]_Q).\eqno(22)$$

Our next task is to consider a system of two $Q$-oscillators
and construct a Hamiltonian having the su$_Q$(2) symmetry.
Using two mutually commuting sets of $Q$-boson operators
($b_1$, $b_1^\dagger$, $N_1$) and ($b_2$, $b_2^\dagger$, $N_2$) one
can construct a Schwinger realization of the su$_Q$(2) generators,
which satisfy the commutation relations $^{34}$
$$ [J_0, J_{\pm}] = \pm J_{\pm}, \qquad J_+ J_- - Q^{-1} J_- J_+ =
[2J_0]_Q, \eqno(23)$$
in the form
$$ J_0=\frac{1}{2} (N_1-N_2), \qquad J_+=(J_-)^\dagger= b_1^\dagger Q^{-N_2/2}
b_2. \eqno(24)$$
The Casimir operator is  $^{34}$
$$ C = Q^{-J_0}([J_0]_Q [J_0+1]_Q + Q^{-1} J_-J_+)
=Q^{-J_0} (J_+J_- + Q [J_0]_Q [J_0-1]_Q) . \eqno(25)$$
This can be written in the form
$$C =- \left[\frac{N_1+N_2}{2}+1\right]_Q \left[-\frac{N_1+N_2}{2}
\right]_Q, \eqno(26)$$
i.e. it depends only on $N_1+N_2$.
One can easily verify that a Hamiltonian of the form
$$H = [N_1 + N_2]_Q, \eqno(27)$$
commutes with the generators of su$_Q$(2) given in Eq. (24), i.e.
has the su$_Q$(2) symmetry.  In order to see the analogy with
Eq. (7) one can write Eq. (27) in the form
$$H = b_1^\dagger b_1 \frac{Q^{N_2}+1}{2} + \frac{Q^{N_1}+1}{2}
 b_2^\dagger b_2. \eqno(28)$$
In order to check the physical content of the coupled $Q$-oscillators
of Eq. (27), we use the Taylor expansion
$$[N]_Q = N +\frac{T}{2} (N^2-N) + \frac{T^2}{12} (2N^3-3N^2+N)
+ \frac{T^3}{24} (N^4-2N^3+N^2) +\dots \eqno(29)$$
Keeping terms up to first order in $T$, the Hamiltonian of Eq. (27)
can be written as
$$H = N_1 \left(1-\frac{T}{2} \right) +\frac{T}{2} N_1^2 +
      N_2 \left(1-\frac{T}{2} \right) +\frac{T}{2} N_2^2 + T N_1N_2 +\dots
\eqno(30)$$
The first two terms in this expansion describe an anharmonic
oscillator of the Morse type with self-anharmonicity $^{35}$
$$x= \frac{T/2}{1-T/2}.\eqno(31)$$
The third and fourth terms in Eq. (30) describe another Morse
oscillator with the same self-anharmonicity. The last term in Eq. (30)
describes the cross-anharmonicity between the two Morse oscillators,
imposed by the overall su$_Q$(2) symmetry. The cross-anharmonicity is
of the same type as the one encountered in the empirical Eqs
(1) and (2).

We conclude therefore that the system of two $Q$-oscillators coupled
so that the total system has the su$_Q$(2) symmetry is equivalent
to  the system of two coupled Morse oscillators usually used
for the description of vibrational spectra of triatomic molecules.

The fact that $Q$-oscillators turn out to be suitable for the description 
of molecules, while $q$-oscillators do not, can be traced back to the 
physical content of these oscillators. It has been found that the 
$Q$-oscillators correspond to WKB equivalent potentials which are harmonic
oscillator potentials with $x^4$ perturbations$^{36}$, while the 
$q$-oscillators
correspond to WKB equivalent potentials which are harmonic oscillator 
potentials with $x^6$ perturbations$^{12}$. 
The relation between the $Q$-oscillators 
and the Morse potential (which is also a harmonic oscillator potential with
leading $x^4$ perturbations) has also been clarified $^{37}$. 

 This construction can be easily generalized to the case of $l$
coupled $Q$-oscillators, having the su$_Q$($l$) overall symmetry.
The Hamiltonian then reads
$$ H = [N_1 + N_2 + \dots + N_l]_Q.\eqno(32)$$
In order to discuss the physical content of this Hamiltonian,
we write explicitly the result for the case with $l=3$, using
the Taylor expansion of Eq. (29) and keeping terms only up to
first order in $T$
$$H(N_1,N_2,N_3)= \sum_{i=1}^3\left(\left(1-\frac{T}{2}\right) N_i + 
\frac{T}{2}
N_i^2\right) + T(N_1 N_2+ N_2 N_3 + N_3 N_1). \eqno(33)$$
We remark that the result is three identical oscillators with
self-an\-ha\-rmo\-ni\-ci\-ty given by Eq. (31) and coupled by cross-anharmonic
terms of the type $N_i N_j$ multiplied by the same strength $T$. It is worth
recalling that this situation is realized in highly symmetric
molecules, such as benzene $^{21-23}$ and the octahedral XY$_6$
molecules treated in $^{20}$. In these molecules all bonds are equivalent and 
all diagonal cross-an\-ha\-rmo\-ni\-ci\-ties among them have the same
strength, as implied by the su$_Q$($l$) (su$_Q$(3)) symmetry in Eq. (32)
(Eq. (33)).

In Eq. (30) it is clear that the cross-anharmonicity strength $T$ and
the self-anharmonicity of the oscillators $x$ are connected as described
by Eq. (31). This connection is imposed by the symmetry used. Such constraints
(imposed by symmetries) 
between the parameters can be useful in reducing the total number of 
free parameters in a model, as pointed out in $^{38}$. Similar constraints
appear in the case of the vibron model, where they have been  found 
particularly useful in reducing the total number of free parameters when 
dealing with four-atomic or larger molecules $^{39}$. In order to have 
larger flexibility, though, one might wish to avoid the connection 
imposed by Eq. (31).  This can be achieved through the use of 
generalized deformed oscillators $^{11}$ (deformed anharmonic oscillators)
of the form
$$ H_i= [N_i + c_i N_i^2]_Q. \eqno(34)$$
(Such oscillators have already been successfully used  for the
description of vibrational spectra of diatomic molecules $^{11}$.)
Two of these oscillators can be coupled in a way similar to that
described by  Eq. (28), giving
$$ H (N_1,N_2) = [N_1 + c_1 N_1^2]_Q \frac{Q^{N_2+c_2 N_2^2}+1}{2}+
\frac{Q^{N_1+c_1 N_1^2}+1}{2} [N_2+c_2 N_2^2]_Q $$ $$=
[N_1 +c_1 N_1^2 + N_2 +c_2 N_2^2]_Q.\eqno(35)$$
Using the Taylor expansion of Eq. (29) and keeping terms only up to
first order in $T$, we have
$$H (N_1, N_2)= (N_1 + c_1 N_1^2) \left(1-\frac{T}{2}\right) +
\frac{T}{2} (N_1^2 + 2 c_1 N_1^3 + c_1^2 N_1^4) $$ $$+
(N_2 + c_2 N_2^2) \left(1-\frac{T}{2}\right) +\frac{T}{2} (N_2^2 +2 c_2 N_2^3
+ c_2^2 N_2^4) $$ $$+ T ( N_1 N_2 + c_1 N_1^2 N_2 + c_2 N_1 N_2^2
+ c_1 c_2 N_1^2 N_2^2).\eqno(36)$$
The physical content of this equation is clear. The first two terms
describe an oscillator with self-anharmonicity (defined as the ratio
of the coefficient of the $N_i^2$ term over that of the $N_i$ term)
$$x_1 = c_1 + \frac{T/2}{1-T/2}, \eqno(37)$$
the third and fourth terms describe an oscillator with
self-anharmonicity
$$x_2 = c_2 +\frac{T/2}{1-T/2}, \eqno(38)$$
while the leading cross-anharmonicity between the two oscillators,
contained in the fifth term, is $T N_1 N_2$.
Therefore Eq. (36) describes to lowest order two Morse oscillators
coupled by the lowest order cross-anharmonicity usally used empirically.

Several comments are now in place:

i) From comparisons to experimental data of diatomic molecules it
is known $^{11}$ that $T$ and $c_i$ are small. Therefore
their present use as small parameters is justified. In particular
it is known that in general $c_i N_i << 1$, so that the terms cubic and 
quartic in $N_i$ occuring in Eq. (36) are of magnitude smaller than the 
corresponding leading terms which are quadratic in $N_i$. 

ii) In Eq. (36) the connection between the self-anharmonicity constants
and the cross-anharmonicity strength is destroyed, as wished. The
$Q$-de\-form\-ation is connected to the cross-anharmonicity strength $T$,
while its effect on the self-anharmonicity constants, seen in Eqs. (37)
and (38), is a simple renormalization.

iii) In Eq. (36) the two oscillators are characterized by different
self-anharmonicities, given by Eqs (37) and (38). This is a feature needed in 
several molecules in which one has to treat inequivalent vibrational 
modes. It is clear that oscillators with different self-anharmonicities occur
only in the case of the generalized deformed oscillators used in Eq. (36)
and not in the case of the $Q$-oscillators used in Eq. (30).  

iv)  The generalization of Eq. (35) to $l$ oscillators reads
$$H(N_1, N_2, \dots, N_l)= [N_1 +c_1 N_1^2 + N_2 +c_2 N_2^2
+\dots + N_l + c_l N_l^2]_Q.\eqno(39)$$
Using the Taylor expansion of Eq. (29) it is clear that to lowest order
$l$ Morse oscillators with self-anharmonicities similar to the ones of Eqs
(37), (38) are obtained, while the cross-anharmonicities among them are of the
$T N_i N_j$ type.

v) In highly symmetric molecules, such as benzene $^{21-23}$
and the octahedral XY$_6$ molecules used in $^{20}$, all bonds are
equivalent. Therefore in Eq. (39) one should use $c_1=c_2=\dots c_l$.
In these molecules the diagonal cross-anharmonicities among the various
bonds are also of the same type $^{20-23}$, therefore the
occurence of the common coefficient $T$ in front of the cross-anharmonic 
terms $N_i N_j$ is justified.

vi) Eq. (35) clearly shows that comultiplication is absent in this
extended model, even in the case with $c_1=c_2$. (For comultiplication
to be present, one should have obtained $c (N_1+N_2)^2$ instead of
$c N_1^2 +c N_2^2$ in the last term of Eq. (35).) Thus the Hopf algebraic 
structure, which is present in the Hamiltonians of Eqs (27) and (32),
is lost in the case of the Hamiltonian of Eq. (35). 

vii) In order to obtain better agreement with the data, one has to add to
the Hamiltonians of Eqs. (1), (2) non-diagonal interaction terms of the form
$^{17,18}$
$$ H_{ij} = a_i^\dagger a_j + a_j^\dagger a_i.\eqno(40)$$
Such terms are also used by the Iachello and Oss approach $^{20-23,40}$. 
Alternatively, terms introducing Darling--Dennison couplings $^{16}$
$$ H_{ij}^{DD}= a_i^\dagger a_i^\dagger a_j a_j + a_j^\dagger a_j^\dagger
a_i a_i \eqno(41)$$
can be used $^{19,40}$. It is clear that such terms are not contained 
in the Hamiltonians of Eqs (27), (32), (35), (39) considered here. 
The matrix elements of such operators can be calculated in a 
straightforward way. For example, the matrix elements of a  term 
$b_i^\dagger b_j$ added to the Hamiltonian of Eq. (27) are
$$ < n_i+1, n_j-1 | b_i^\dagger b_j | n_i , n_j>= \sqrt{ [n_i+1]_Q [n_j]_Q},
\eqno(42)$$
because of Eq. (20). Similarly one has 
$$ <n_i+2, n_j-2| b_i^\dagger b_i^\dagger b_j b_j | n_i,  n_j> =
\sqrt{ [n_i+2]_Q [n_i+1]_Q [n_j]_Q [n_j-1]_Q }.\eqno(43)$$


viii) An algebraic model based on two coupled anharmonic vibrations for 
the description of ABA triatomics has been developed in $^{41}$. The 
diagonal part of its Hamiltonian can be rewritten in the form of Eq. (2),
 involving  three free parameters. To leading order it is then equivalent 
to Eq. (36) of the present formalism, which contains the same leading 
terms and involves three free parameters as well. 

In summary, we have proved that a system of $n$ $Q$-deformed
oscillators coupled so that the total Hamiltonian is characterized
by the su$_Q$(n) symmetry is equivalent, to lowest order
approximation, to a system of $n$ identical Morse oscillators coupled
by the cross-anharmonicity usually used empirically in the description
of vibrational spectra of polyatomic molecules.  This symmetry
also implies a relation between the self-anharmonicity of the Morse
oscillators and the cross-anharmonicity strength, which can be avoided by
 replacing the $Q$-oscillators by deformed anharmonic
oscillators (generalized deformed oscillators), 
at the cost of losing the Hopf algebraic structure.
No Darling--Dennison or other nondiagonal interaction 
terms are contained in this scheme.
For introducing such terms in the quantum algebraic framework the approach 
of $^{40}$ can be used. 

{\bf Acknowledgement}

One of the authors (DB) has been supported by the EU under contract 
ERBCHBGCT930467.

\newpage

\parindent=0pt
\centerline{\bf References}
\bigskip

\ref $^1$ V. G. Drinfeld, in {\it Proceedings of the
International Congress of Mathematicians}, edited by  A. M. Gleason
(American Mathematical Society, Providence, RI, 1986) p. 798.

\ref $^2$ M. Jimbo, Lett. Math. Phys.  {\bf 11}, 247 (1986).

\ref $^3$ E. Abe, {\it Hopf Algebras} (Cambridge U. Press, Cambridge, 1977)

\ref $^4$ M. Jimbo, in {\it Braid Group, Knot Theory and
Statistical Mechanics}, edited by C. N. Yang and M. L. Ge
(World Scientific, Singapore, 1989) p. 111.

\ref $^5$ L. C. Biedenharn, J. Phys. A {\bf 22}, L873 (1989).

\ref $^6$ A. J. Macfarlane, J. Phys. A  {\bf 22}, 4581 (1989).

\ref $^7$ C. P. Sun and H. C. Fu, J. Phys. A {\bf 22}, L983 (1989). 

\ref $^8$ D. Bonatsos, P. P. Raychev, R. P. Roussev and Yu. F. Smirnov,
 Chem. Phys. Lett. {\bf 175}, 300 (1990).

\ref $^9$ Z. Chang and H. Yan, Phys. Lett. A {\bf 154}, 254 (1991).

\ref $^{10}$ Z. Chang, H. Y. Guo and H. Yan, Commun. Theor. Phys. {\bf 17}, 183
(1992). 

\ref $^{11}$ D. Bonatsos and C. Daskaloyannis, Phys. Rev. A {\bf 46}, 75 
(1991).

\ref $^{12}$ D. Bonatsos, C. Daskaloyannis and K. Kokkotas,
 J. Math. Phys. {\bf 33}, 2958 (1992).

\ref $^{13}$ Y. Alhassid, F. G\"ursey and F. Iachello, 
Ann. Phys. {\bf 148}, 346 (1983).

\ref $^{14}$ P. P. Raychev, Adv. Quant. Chem.  {\bf 26}, 239 (1995). 

\ref $^{15}$ G. Herzberg, {\it Molecular Spectra and Molecular
Structure} Vol. III {\it Electronic Spectra and Electronic Structure
of Polyatomic Molecules} (Van Nostrand, Toronto, 1979).

\ref $^{16}$ B. T. Darling and D. M. Dennison, Phys. Rev. {\bf 57}, 128 (1940).

\ref $^{17}$ M. E. Kellman, J. Chem. Phys. {\bf 81}, 389 (1984).

\ref $^{18}$ M. E. Kellman, Chem. Phys. Lett. {\bf 113}, 489 (1985).

\ref $^{19}$ M. E. Kellman, J. Chem. Phys. {\bf 83}, 3843 (1985).

\ref $^{20}$ F. Iachello and S. Oss, Phys. Rev. Lett. {\bf 66}, 2976 (1991).

\ref $^{21}$ F. Iachello and S. Oss, Chem. Phys. Lett. {\bf 187}, 500 (1991).

\ref $^{22}$ F. Iachello and S. Oss, J. Mol. Spectrosc. {\bf 153}, 225 (1992).

\ref $^{23}$ F. Iachello and S. Oss, J. Chem. Phys. {\bf 99}, 7337 (1993). 

\ref $^{24}$ F. Iachello and R. D. Levine, J. Chem. Phys. {\bf 77}, 3046
(1982).

\ref $^{25}$ O. S. van Roosmalen, F. Iachello, R. D. Levine and
A. E. L. Dieperink, J. Chem. Phys. {\bf 79}, 2515 (1983).

\ref $^{26}$ P. P. Kulish, in {\it Group Theoretical Methods in Physics}, 
edited by 
V. V. Dodonov and  V. I. Man'ko (Springer-Verlag, Berlin, 1991) p. 195. 

\ref $^{27}$ E. G. Floratos, J. Phys. A {\bf 24}, 4739 (1991).

\ref $^{28}$ C. P. Sun, X. F. Liu, J. F. Lu and M. L. Ge, 
J. Phys. A {\bf 25}, L35 (1992).

\ref $^{29}$ S. Fl\"ugge, {\it Practical Quantum Mechanics} 
(Springer-Verlag, Berlin, 1974).

\ref $^{30}$ M. Arik M and D. D. Coon, J. Math. Phys. {\bf 17}, 524 (1976).

\ref $^{31}$ M. V. Kuryshkin, Annales de la Fondation Louis de Broglie 
{\bf 5}, 111 (1980).

\ref $^{32}$ A. Jannussis, in {\it Hadronic Mechanics and Non-potential
 Interactions}, ed\-i\-t\-ed by H. C. Muyng (Nova Science, Commack, NY, 1991).

\ref $^{33}$ P. P. Kulish and E. V. Damaskinsky, J. Phys. A {\bf 23}, L415
 (1990).

\ref $^{34}$ R. Chakrabarti and R. Jagannathan, J. Phys. A {\bf 24}, L711 
(1991).

\ref $^{35}$ I. L. Cooper, Chem. Phys. {\bf 112}, 67 (1987).

\ref $^{36}$ Th. Ioannidou, Diploma Thesis, U. of Thessaloniki (1993). 

\ref $^{37}$ D. Bonatsos and C. Daskaloyannis, Chem. Phys. Lett. {\bf 203},
150 (1993). 

\ref $^{38}$ M. E. Kellman, Annu. Rev. Phys. Chem. {\bf 46}, 395 (1995). 

\ref $^{39}$ F. Iachello, S. Oss and R. Lemus, J. Mol. Spectrosc. {\bf 149},
132 (1991).  

\ref $^{40}$ D. Bonatsos and C. Daskaloyannis, Phys. Rev. A {\bf 48}, 3611
 (1993).  

\ref $^{41}$ O. S. van Roosmalen, I. Benjamin and R. D. Levine, J. Chem.
Phys. {\bf 81}, 5986 (1984). 

\end{document}